\documentclass[pra,aps,amsmath,floatfix,amssymb,twocolumn]{revtex4-2}
\usepackage[colorlinks,bookmarks=false,citecolor=blue,linkcolor=blue,urlcolor=blue]{hyperref}
\usepackage[all]{hypcap}   
\usepackage{amsmath,amssymb,orcidlink,float,nicefrac,svg,placeins,soul,ulem,graphicx,dcolumn,bm,verbatim,color,flafter,hhline,physics,hyperref}
\graphicspath{{figures/}{/}}
\usepackage[capitalise,compress]{cleveref}
\crefname{section}{Sec.}{Secs.}
\Crefname{section}{Section}{Sections}

\usepackage{xr-hyper}
\begin{document}

\title{\textbf{Localization with Hopping Disorder in a Quasiperiodic Synthetic Momentum Lattice}}

\author{Joel M. Sunil\orcidlink{0009-0008-5807-7883}}
\altaffiliation{These authors contributed equally to this work}
\altaffiliation{Email: \href{mailto:joel.sunil@students.iiserpune.ac.in}{joel.sunil@students.iiserpune.ac.in}, \href{mailto:bharathikannan1130@gmail.com}{bharathikannan1130@gmail.com}}
\affiliation{Indian Institute of Science Education and Research, Pune 411008, India}

\author{J. Bharathi Kannan\orcidlink{0000-0002-1449-684X}}
\altaffiliation{These authors contributed equally to this work}
\altaffiliation{Email: \href{mailto:joel.sunil@students.iiserpune.ac.in}{joel.sunil@students.iiserpune.ac.in}, \href{mailto:bharathikannan1130@gmail.com}{bharathikannan1130@gmail.com}}
\affiliation{Indian Institute of Science Education and Research, Pune 411008, India}

\author{Monu Bhartiya \orcidlink{0009-0005-8487-4585}}
\affiliation{Indian Institute of Science Education and Research, Pune 411008, India}

\author{Rayees A S}
\affiliation{Indian Institute of Science Education and Research, Pune 411008, India}

\author{Shuvarati Roy}
\affiliation{Indian Institute of Science Education and Research, Pune 411008, India}

\author{G. J. Sreejith \orcidlink{0000-0002-2068-1670}}
\affiliation{Indian Institute of Science Education and Research, Pune 411008, India}

\author{M. S. Santhanam \orcidlink{0000-0002-7213-8346}}
\affiliation{Indian Institute of Science Education and Research, Pune 411008, India}

\author{Umakant D. Rapol
\orcidlink{0000-0003-0340-6017}}
\affiliation{Indian Institute of Science Education and Research, Pune 411008, India}

\date{\today}

\begin{abstract}
Quasiperiodicity is easily realized with ultracold atoms in optical lattices and has been used to simulate the delocalization-localization transition. Random local potentials can also be implemented by spatial light modulators. 
Programmable hopping disorder, however, remains difficult to realize in conventional optical lattices. Momentum-space lattice (MSL) offers a platform to address this.
We implement the MSL using an ${^{87}{\text{Rb}}}$ Bose-Einstein condensate via multiple Bragg diffraction and benchmark the platform by simulating the dynamics of a wavepacket in the presence of hopping disorder on top of the generalized Aubry Andr\'e (GAA) model. The system tests the effect of the programmable hopping disorder when added to an otherwise localized/delocalized background tunable by the Aubry Andr\'e potential strength as well as in systems with mobility edges.
The uncorrelated hopping disorder smooths the AA transition into a crossover between weakly (large localization lengths) and strongly localized regimes; the generic enhancement in localization is reflected in the suppressed wavepacket spread.
We also implement spatially correlated hopping disorder and compare its dynamics with uncorrelated disorder of the same final width, demonstrating a correlation-induced enhancement in spreading. Measured dynamics agrees with trends found in the ideal tight-binding simulations. Trends in non-universal features that depend on the microscopic structures and the short term dynamics as a function of quasiperiodic potential strength, hopping-disorder strength, and GAA deformation also match with the numerical results; with better agreement at larger hopping-disorder strengths.
Numerical simulations of the momentum-space lattice dynamics provide insights into sources of corrections. 
Our results demonstrate that MSL provides a versatile platform for realizing more general disordered quantum systems with programmable hopping amplitudes, beyond quasiperiodic models.
\end{abstract}

\maketitle

\begingroup
\renewcommand\thefootnote{*}
\footnotetext{These authors contributed equally to this work.}
\endgroup

\section{Introduction}
Anderson localization, the suppression of quantum transport by disorder-induced interference
\cite{anderson1958absence,evers2008anderson,lagendijk2009fifty}, is a universal wave phenomenon across different physical platforms. In ultracold atoms, Anderson localization has been observed for matter waves in controlled disorder \cite{billy2008direct,roati2008anderson,kondov2015disorder,semeghini2015measurement}. In photonic systems, localization has been studied in disordered media and lattices \cite{wiersma1997localization,schwartz2007transport,lahini2009observation}, while related signatures have also been explored through statistical measurements of photon localization \cite{chabanov2000statistical}.
The scaling theory of localization established that, for non-interacting particles with uncorrelated disorder, all states are localized in one- and two-dimensions, whereas in three dimensions a disorder-driven quantum phase transition separates extended and localized phases via an energy-dependent mobility edge \cite{abrahams1979scaling}. The one-dimensional Aubry–Andr\'e (AA) model \cite{aubry1980analyticity,harper1955single} provides a complementary route to localization via deterministic quasiperiodic potentials and exhibits a sharp self-duality–protected transition.

Ultracold atoms have enabled the exploration of quasiperiodic localization across increasingly rich settings, from single particle \cite{roati2008anderson} to interacting and periodically driven many-body systems \cite{schreiber2015observation,bordia2017periodically}, including mobility-edge physics in generalized Aubry-Andr\'e models \cite{luschen2018SPME,an2021interactions}. Photonic lattices \cite{lahini2009observation,kraus2012topological,verbin2013observation} and superconducting circuits \cite{li2023observation,guo2021stark} have offered complementary settings to study localization and related phenomena in quasiperiodic systems. Specific deformations of the AA model with an energy-dependent duality give rise to the generalized AA (GAA) class \cite{biddle2010predicted,ganeshan2015nearest}, which hosts richer localization phenomena, including energy-dependent mobility edges~\cite{li2017mobility,li2015many,roy2018multifractality}. Variants of these quasiperiodic models have now been realized in synthetic quantum platforms, underscoring the broad relevance of quasiperiodic localization.

However models with {\it off-diagonal} (hopping) disorder remain experimentally unexplored due to insufficient local control in conventional platforms \cite{shimasaki2024anomalous,luschen2018SPME,bordia2017periodically}. We use a Momentum Space Lattice (MSL) as an analog simulation platform with tunable hopping disorder. We implement the hopping disorder in the background formed by a GAA model which can be tuned from extended to localized regimes.
Uncorrelated hopping disorder should suppress transport generically~\cite{roy2020interplay}. Correlated hopping disorder is also expected to localize the delocalized phase of the AA model (though localization lengths typically are larger when the correlations are stronger).  In the localized phase, correlations can have the opposite effect in certain regions at least as we show using numerical analysis in Sec.~\ref{sec:model}. Here we use a synthetic momentum-space lattice (MSL) realized by coherently coupling discrete momentum states of an $^{87}$Rb Bose–Einstein condensate using far-detuned laser fields \cite{meier2016simulator}. 
Unlike conventional real-space lattices, the MSL provides a scalable approach to site-resolved control over every tunneling amplitude, enabling a direct implementation of arbitrary hopping disorder with tunable spatial correlations. We note that the GAA model has been realized in an MSL experiment \cite{an2021interactions}, with evolution times similar to what is achieved here. 
We benchmark the MSL by simulating the wavepacket dynamics starting from a state at the center of the 21-site lattice and compare it with (i) ideal tight-binding model as well as (ii) numerical simulations of the MSL at a non-interacting level. We find that MSL successfully captures the trends in short time dynamics of wavepacket spread as a function of a broad set of parameters including the hopping disorder strength $\sigma$, GAA deformation $\alpha$ and GAA local potential strength $\lambda$. 
We also study hopping disorder correlations and show that statistically the delocalizing effect of correlations is captured correctly by the simulator. 

The paper is organized as follows. Section~\ref{sec:model} introduces the generalized Aubry–André model with hopping disorder, defines the localization diagnostics, defines the disorder distributions and examines the effects of disorder strength and spatial correlations on its eigenstates using numerical simulations of the tight binding model. We also consider the properties of wavepacket spread and transmission as a function of correlations with the goal of understanding the expected behavior in the MSL simulations. Section~\ref{sec:experiment} describes the implementation of the model in a 21-site momentum-space lattice and the simulations used to account for experimental effects. Sec.~\ref{sec:Results} compares the measured wave-packet dynamics with ideal tight-binding and numerical momentum-space-lattice simulations for both uncorrelated hopping disorder and spatially correlated disorder across parameter regimes; and we conclude in Sec.~\ref{sec:Summary}. The appendices provide details of the numerical simulations and transmission calculations, and additional inverse participation ratio data from the measurements.

\section{Model\label{sec:model}}
\begin{figure}[htbp]
    \centering
    \includegraphics[width=0.97\linewidth]{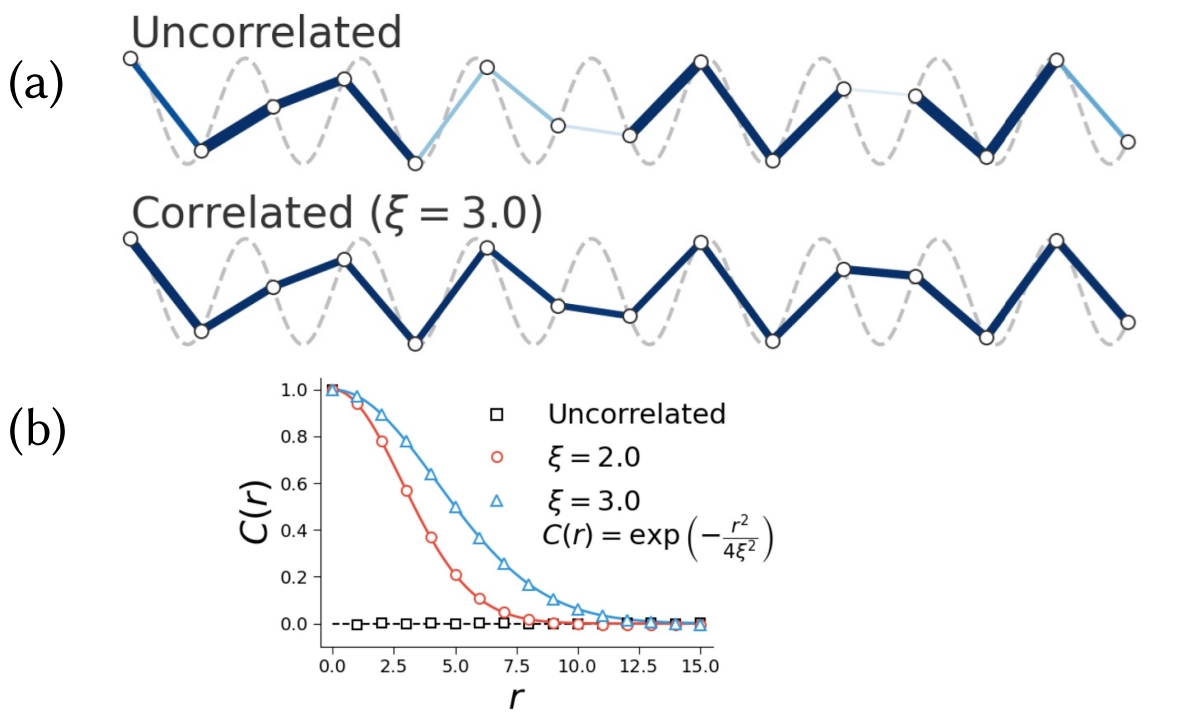}
    \includegraphics[width=0.97\linewidth]{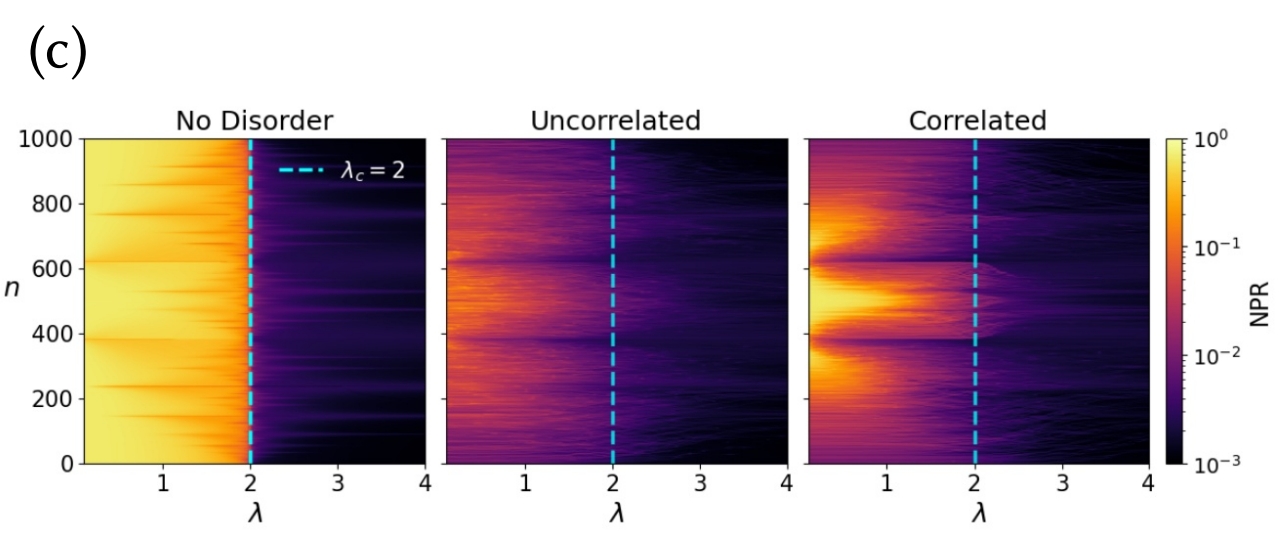}
    \caption{(a) Schematic depiction of AA chain with hopping disorder, circles denote lattice sites, the dashed curve the potential and bond thickness represents the hopping strength.  Correlated disorder yields a smoother spatial dependence of hopping. (b) Autocorrelation of the hopping disorder. (c) NPR of eigenstates for a system size $L{=}1000$ as a function of $\lambda$ and eigenstate index $n$, shown for the AA system (left), AA with uncorrelated hopping disorder (center), and correlated hopping disorder with $\xi=3$ (right). The vertical dashed line marks the self-dual critical point $\lambda_c=2$, as a guide to the eye.}
    \label{fig:IPR_lambda_sigma}
\end{figure}
We consider a 1D GAA chain with hopping disorder, described by the Hamiltonian
\begin{equation}
H{=}\sum_{i=-n/2}^{n/2-1}t_{i}({c_{i+1}^{\dagger}c_{i}}{+}\text{h.c.}) {+} \lambda\sum_{i=-n/2}^{n/2}\frac{c_{i}^{\dagger}c_{i}\cos(2\pi\beta i{+}\phi)}{1{-}\alpha\cos(2\pi\beta i{+}\phi)}
\label{eq:gaa_hamiltonian}
\end{equation}
where $c_i^\dagger$ creates a particle in synthetic lattice site $i$, $t_i$ is the nearest-neighbor hopping amplitude between sites $i$ and $i+1$, $\lambda$ sets the quasiperiodic potential strength, $\alpha$ parameterizes the GAA deformation~\cite{biddle2010predicted,ganeshan2015nearest}. $L$ (=21 in our experiment) is the system size and $n=L-1$ is the number of hopping bonds for open boundary condition. The incommensurate ratio $\beta{=}{(\sqrt{5}-1)}/{2}$ ensures quasiperiodicity, and $\phi$ is a phase offset. We set $\phi=0$ throughout the paper. The onsite energy is then given by
\begin{equation}
    \mu_{i}=\lambda\frac{\cos(2\pi\beta i)}{1{-}\alpha\cos(2\pi\beta i)}\:.
\end{equation}
We impose open boundary conditions throughout. In the clean limit with $\alpha{=}0$ and uniform hopping $t$, Eq.~\ref{eq:gaa_hamiltonian} reduces to the standard AA model, which exhibits a self-duality-protected localization transition at $\lambda_c{=}2t$ with all eigenstates simultaneously transitioning from extended to localized as $\lambda$ increases~\cite{aubry1980analyticity,harper1955single}. The generalization with $\alpha{\neq}0$ creates a model with an energy dependent self-duality and can support mobility edges~\cite{roy2018multifractality, li2017mobility}.

Ref.~\cite{roy2020interplay} studied an AA chain with random hopping disorder of width $\sigma$ around $t=1$. They find that hopping disorder broadens the clean localization transition at ($\lambda_c=2$) as expected, producing eigenstates with energy-dependent localization. For AA strength smaller than $\lambda_c$, stronger hopping disorder generally increases localization. Above the transition, localization is generally enhanced by hopping disorder, except near the spectral edges, where a reduction is observed. 

\begin{figure*}
    \includegraphics[width=\textwidth]{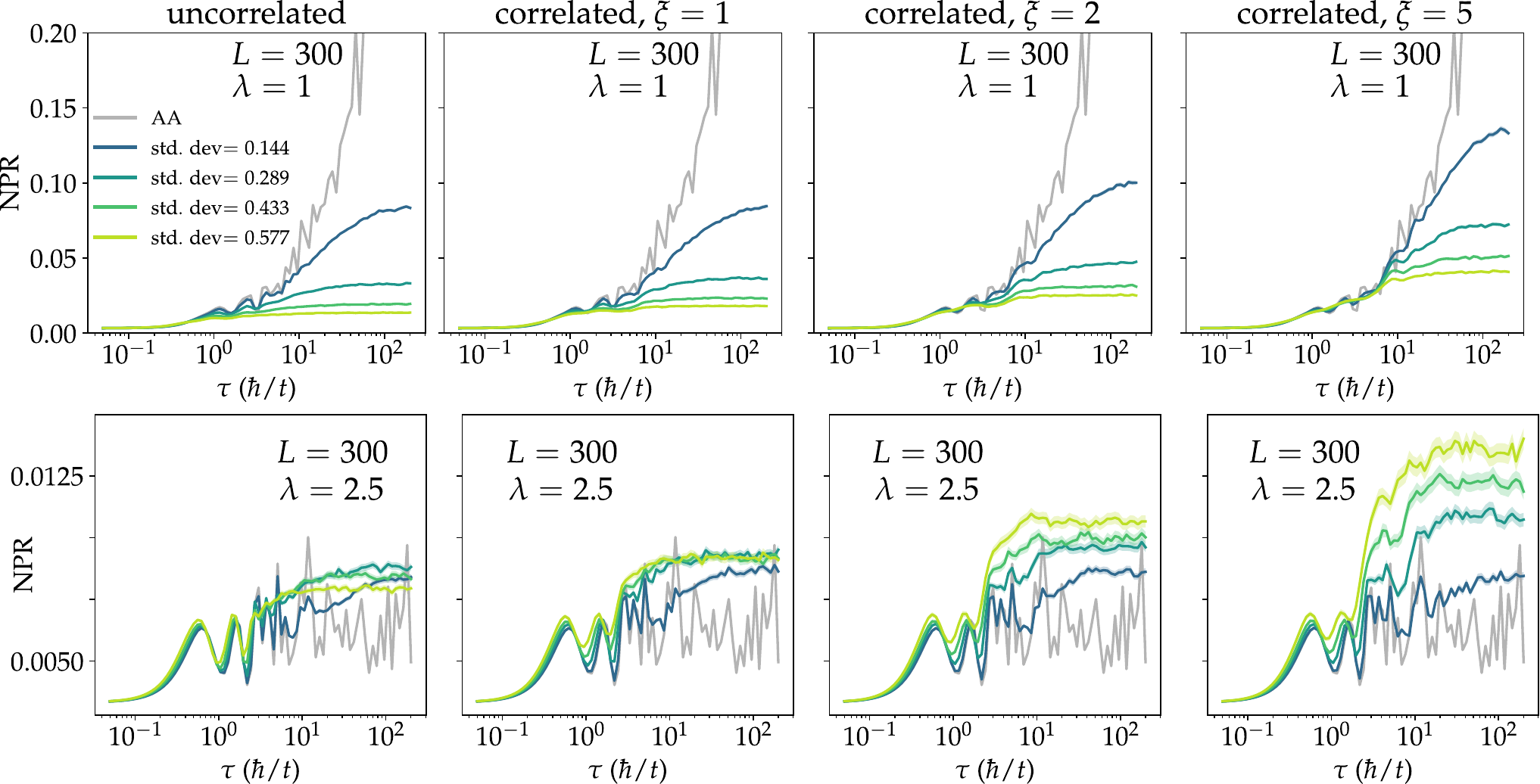}
    \caption{Time evolution of the NPR in the AA model but with hopping disorder of different variances. Different panels show the dynamics for different disorder correlation length scales $\xi$ obtained by smoothening the uncorrelated disorder profiles. Top row shows the results when disorder is added to the delocalized AA phase. Bottom row shows the same but for disorder added to the localized AA phase. Results indicate suppressed wavefunction when disorder strength increases. Even when the disorder width/variance is kept fixed, smoothening results in increased wavefunction spread. Results here show NPR dynamics in a long chain of size $L=300$. Qualitatively similar results are found in the case of $L=21$ also (See Appendix). NPR values are averaged over 1000 realizations.\label{fig:NPRDynamics}}
    
\end{figure*}

We characterize localization using measures accessible in the experiment. The normalized participation ratio (NPR) for a state $\vert \psi \rangle$ is defined as~\cite{evers2008anderson} $\mathrm{NPR} = 1/({L}\sum_i |\langle i|\psi\rangle|^4)$.
 In experiments, eigenstates are not directly accessible; instead, we infer localization from the time evolved state $|\psi(\tau)\rangle$ at time $\tau$ starting from an initial state $|\psi(0)\rangle$ localized at the center of the lattice. The NPR is calculated from the measured site-resolved populations (after normalization) $\mathcal{P}_i =|\langle i|\psi(\tau)\rangle|^2$ as
 \begin{equation}
     \mathrm{NPR(\tau)} = \frac{1}{{L}\sum_i \mathcal{P}_{i}^{2}}.
 \end{equation}
The NPR is $\sim 1$ for a fully delocalized state and $1/L$ for a fully localized state. To further probe localization, we measure the return probability (RP), which is the probability of finding the particle in the initial site at time $\tau$ given by $\mathrm{RP}(\tau) =\abs{\langle\psi(0)|\psi(\tau)\rangle}^2$. A large, slowly decaying RP indicates strong localization with suppressed transport, while rapid decay signals delocalization and coherent spreading~\cite{torres2017dynamical}. 
The NPR quantifies the spatial spread of the wavepacket, whereas RP quantifies the localization at the initial position, providing  dynamical signatures of localization.

To study the effect of hopping disorder, we consider both uncorrelated and
spatially correlated nearest-neighbor hopping amplitudes $t_i$ on the $L-1$
bonds of the chain. For uncorrelated disorder, each $t_i$ is drawn
independently from the uniform distribution on $[1-\sigma,\,1+\sigma]$, so
that $\sigma$ is the half-width of the distribution and its standard
deviation is $\Sigma_0{=}\sigma/\sqrt{3}$. Correlated disorder with short-range
correlations is generated by Gaussian smoothing of such an uncorrelated
sequence $\eta_j$,
\begin{equation}
    t_i=\frac{\sum_{j=1}^{L-1} g(d_{ij})\,\eta_j}
             {\sum_{j=1}^{L-1} g(d_{ij})},
    \qquad
    g(r)=e^{-\frac{r^2}{2\xi^2}},
    \label{eq:gaussian_smoothing}
\end{equation}
where $\xi$ sets the correlation length scale and $d_{ij}{=}\min\left(|i-j|,\,L-1-|i-j|\right)$ is the distance between bonds $i$ and $j$ with periodic wrapping. The periodic distance is used only to avoid edge effects in the smoothing; the chain itself retains open boundary conditions. Because the weights in Eq.~\eqref{eq:gaussian_smoothing} sum to unity, smoothing leaves the disorder-averaged hopping at $\bar t=1$ and only suppresses fluctuations about it.

We characterize the resulting spatial correlations in each realization by the normalized
correlation function
\begin{equation}
    C(r)=\frac{\left\langle (t_i-\bar t)(t_{i+r}-\bar t)\right\rangle_{i,\mathrm{dis}}}
              {\left\langle (t_i-\bar t)^2\right\rangle_{i,\mathrm{dis}}},
    \label{eq:correlation_function}
\end{equation}
where $\langle\cdot\rangle_{i,\mathrm{dis}}$ denotes an average over all bond
pairs $(i,\,i+r)$ within the chain, $1\le i\le L-1-r$, and over disorder
realizations. For an infinite chain with $\xi\gg1$, Gaussian smoothing gives
$C(r)\simeq\exp[-r^2/(4\xi^2)]$; for the finite chains considered here, we
compute $C(r)$ directly from the generated realizations.

Smoothing reduces the standard deviation of the hopping amplitudes below $\Sigma_0$. Comparing correlated and uncorrelated disorder at the same $\sigma$ would therefore confound the effect of correlations with that of weaker disorder. To isolate the role of correlations, we compare each correlated realization with its uncorrelated parent sequence $\eta_j$, rescaled about its mean so that its standard deviation matches that of the smoothed sequence. The rescaled sequence remains uncorrelated, so each pair of realizations has the same mean and width and differs only in its spatial correlations. We denote by $\Sigma_\xi$ the standard deviation of the smoothed hopping terms averaged over realizations, which characterizes the typical disorder width at a given $\xi$.

\begin{figure}[htbp]
\includegraphics[width=\columnwidth]{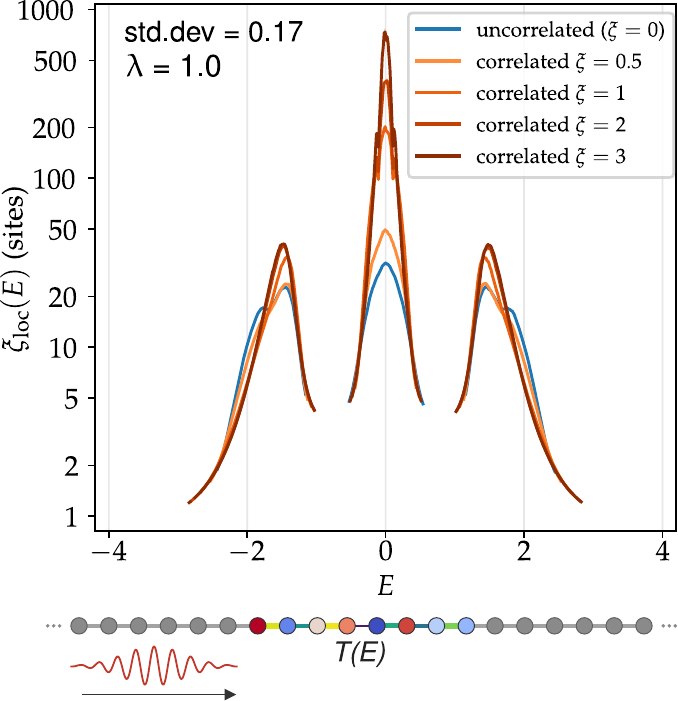}
\caption{Typical localization lengths $\xi_{\rm loc}$ at different energies estimated from calculations of the transmission coefficient $T(E)$ in the AA chain in the delocalized phase with added hopping disorder of lengths up to $L=5000$ connected by a simple infinite lead. Localization length is calculated by fitting the $L$ dependence of $T$ to $e^{-L/\xi_{\rm loc}}$. Details of the calculations can be found in the Appendix. The localization lengths rapidly increase with correlation length scale $\xi$ at most energies. The gaps in the plot represents the energy ranges where the density of states is 0. ~\label{fig:transmission}}
\end{figure}

Before proceeding, we consider the physics expected when hopping disorder is added to the ideal long AA chain. Figure~\ref{fig:IPR_lambda_sigma} shows how hopping disorder and $\xi$ reshape localization in the AA chain ($\alpha=0$) of length $L=1000$. 
Eigenstate-resolved NPR in AA model shows a sharp transition at $\lambda_{\rm c}{=}2$ across which all eigenstates localize. While hopping disorder results in localization of all generic states (with a chiral-symmetry protected divergence in density of states at the band center)~\cite{Eggarter1978}, combining with AA disorder destroys any chiral symmetry, resulting in a clear localization of all eigenstates. Consistent with this expectation, uncorrelated hopping disorder in AA enhances localization (Fig.~\ref{fig:IPR_lambda_sigma}(c)), replacing the transition to a crossover at $\lambda_{\rm cross}{\approx}2$ from a weakly localized to strongly localized regimes. The effect of the disorder can also be seen in the dynamics of the wavepacket spread. Figure~\ref{fig:NPRDynamics} shows the NPR dynamics in the AA model as the disorder variance and correlation length scale $\xi$ are tuned. Addition of uncorrelated disorder and increase in its width results in suppressed wavefunction spread. When the disorder is added to localized phase of the AA model the disorder width dependence of the localization length is weak. Qualitatively similar results are obtained in smaller chains as shown in the Appendix.

In the short-range  correlated disorder case, Gaussian smoothing produces smoothly varying hopping along the lattice [Fig.~\ref{fig:IPR_lambda_sigma}(a)], as characterized by the spatial autocorrelation function [Fig.~\ref{fig:IPR_lambda_sigma}(b)]. 
Correlated disorder in the AA model results in energy-dependent and a wider ranged localization lengths in the $\lambda{<}2$ regime as reflected in the wide fluctuations in the NPR values in right panel of Fig.~\ref{fig:IPR_lambda_sigma}(c). Correlations suppresses the effect of disorder even if the variance of the disorder distribution is held fixed. In the NPR dynamics shown in Fig.~\ref{fig:NPRDynamics} this is reflected in enhanced wavefunction spread as $\xi$ is increased. We understand this as follows, smoothening causes contiguous regions where the hopping strengths are nearly constant thereby suppressing the part of the scattering induced by the hopping disorder. An anomalous behavior is observed in localized phase when the correlations are large (bottom right most panel of Fig.~\ref{fig:NPRDynamics}). At large $\xi$ stronger disorder leads to contiguous regions of strong bonds which may locally appear like an extended phase and can lead to faster spread of wavefunctions. Figure~\ref{fig:transmission} shows the energy dependence of the typical localization lengths $\xi_{\rm loc}$ estimated from transmission calculations at different energies in long chains. Localization lengths diverge in the pure AA model at $\lambda=1$ as it is in delocalized phase. However disorder generates a finite localization length which increases sharply with increasing disorder correlations $\xi$. In the strongly localized regime of AA, the disorder do not cause a similar clear change in localization lengths. 

\section{Implementation of Momentum Space Lattice\label{sec:experiment}}
\begin{figure}[htbp]
    \centering
    \includegraphics[width=0.9\columnwidth]{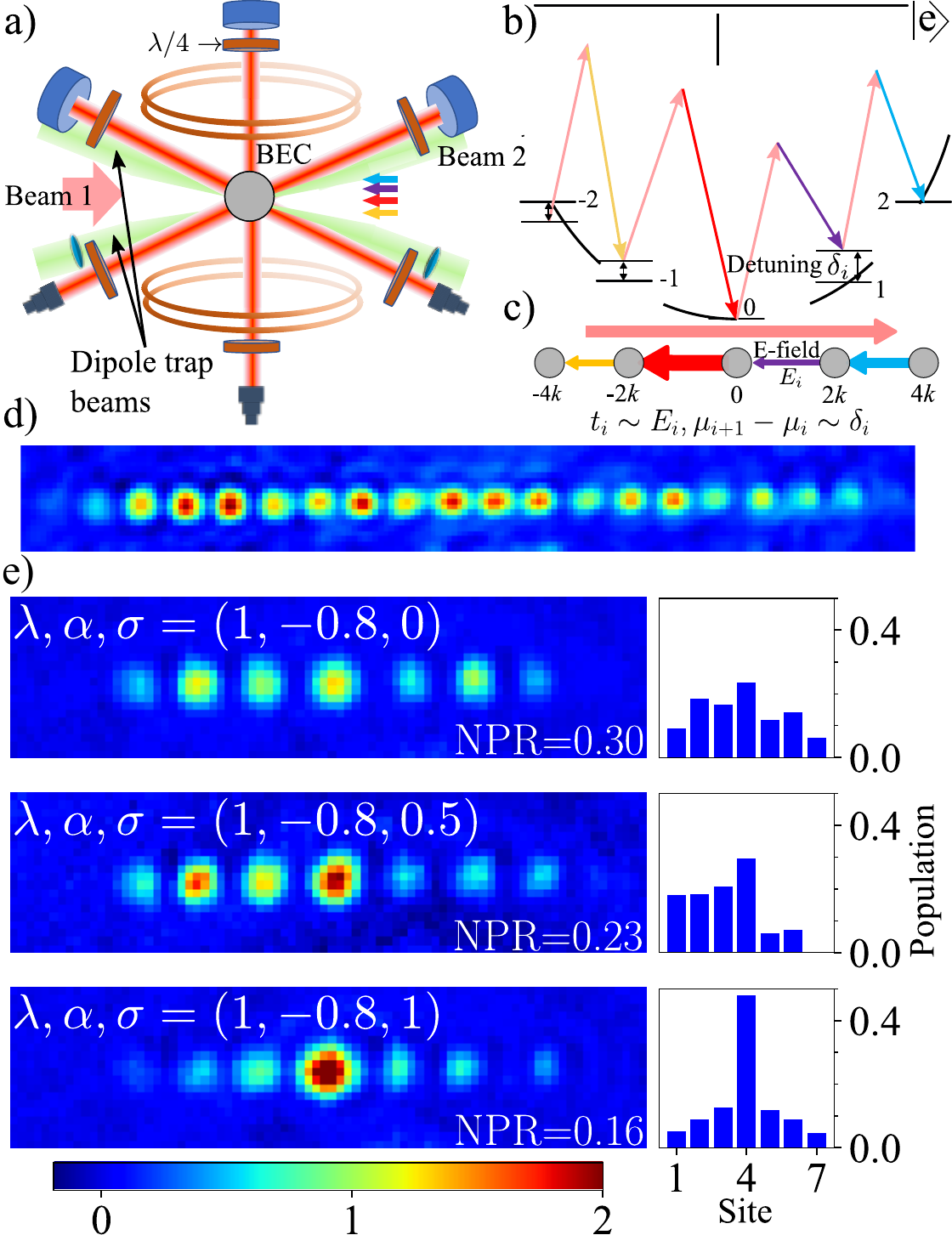}
        \caption{(a) Schematic of the experimental setup. The MOT (initial cooling) followed by the XDT (for evaporative cooling) are used to achieve a BEC. Then, the beam-1 and the multi-tone beam-2 are turned on to create the MSL.  Energy and momentum space pictures of the experiment are shown in (b) and (c), respectively. In (b) the excited intermediate state for the Bragg transitions is labeled by $|e\rangle$. The hopping term ($t_i$) is controlled by the electric-field amplitude $E_i$, while the detuning ($\delta_i$) controls the onsite-energy difference $\mu_{i+1}-\mu_i$ between neighboring sites. (d) Representative optical density (OD) plots (from absorption imaging) for a highly delocalized case: $\lambda, \alpha, \sigma=(0.2, 0, 0.5)$.(e) (left) OD plots of the time-evolved BEC, which show localization with increasing hopping disorder $\sigma$. (right) Estimated population vs position.
    \label{fig: Experimental setup and absorption images}}
\end{figure}
\begin{figure*}[!]
     \centering
     \includegraphics[width=0.88\linewidth]{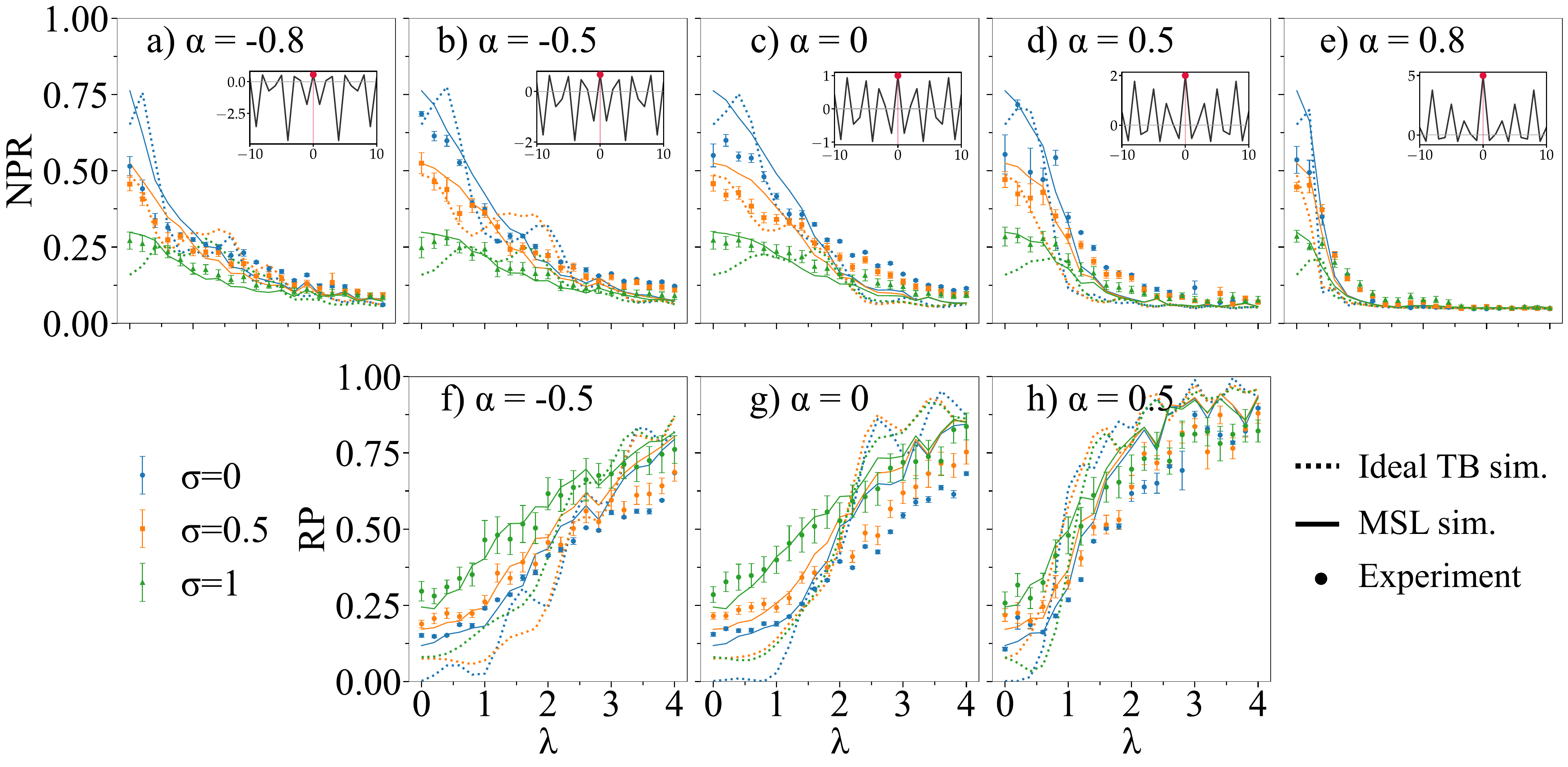}
     \caption{NPR (a-e) and RP (f-h) after time evolution as a function of quasiperiodic potential strength $\lambda$. {The dotted lines are the ideal tight-binding (TB) simulations which do not account for possible experimental deviations such as the momentum width of the BEC, while the solid lines are the simulations of the MSL experiment which consider these deviations. The markers with error bar (standard error) are experimental values. The inset shows the Aubry-Andre potential for the corresponding $\alpha$ values.}}
     \label{fig: exp_NPR_RP_diff_alpha}
\end{figure*}

We realize the GAA model for 21 sites using a $^{87}$Rb BEC (about $4{\times}10^{4}$ atoms) as shown in Fig.~\ref{fig: Experimental setup and absorption images}. Multiple Bragg transitions are coupled using $1064\,\mathrm{nm}$ laser beams to create a lattice in the momentum space similar to Refs.~\cite{meier2016simulator, an2021interactions}. ${}^{87}{\text{Rb}}$ atoms are loaded from a Zeeman slower into a Magneto-Optical Trap (MOT), where they are cooled to a temperature of $50\,\mu\mathrm{K}$. They are then transferred to a crossed dipole trap (XDT) formed by two $1064\,\mathrm{nm}$ laser beams. Further cooling is achieved through forced evaporation in the XDT to $50\,\mathrm{nK}$ to create a BEC. A pair of counter-propagating lasers at $1064\,\mathrm{nm}$ are used to couple multiple Bragg transitions. In a Bragg transition, the atom absorbs light from one laser and undergoes stimulated emission in the direction of the other laser. Its momentum changes by $2\hbar k$, without any internal state change. We use this momentum scattering to simulate nearest neighbor hopping in a lattice of 21 momentum states $p=-20\hbar k, -18\hbar k,\dots20\hbar k$. Each Bragg transition occurs at a unique frequency, allowing independent control of the hopping amplitudes and the energy differences between neighboring synthetic lattice sites. The electric-field amplitude \(E_i\) of the \(i\)th tone controls the hopping amplitude \(t_i\), while its detuning \(\delta_i\) controls the onsite-energy difference \(\mu_{i+1}-\mu_i\) between neighboring sites (Fig.~\ref{fig: Experimental setup and absorption images}(b,c)). 
The required multi-tones are generated by an Arbitrary Waveform Generator and fed to the Acousto-Optic Modulator to control the Bragg laser beams. The Bragg laser beams are pulsed on for 1ms immediately after the evaporation sequence is complete, and the XDT is turned off. We restrict the hopping terms $t_i$ to a maximum of $h\times1.4\,\mathrm{kHz}$, since larger values produce significant off-resonant excitation. Hopping and disorder strengths hereafter are quoted in units of $h\times0.7\:\mathrm{kHz}$.

Figure~\ref{fig: Experimental setup and absorption images}(e) shows the optical absorption images and the estimated site dependent population \textcolor{blue}{$\mathcal{P}_i$} for MSL simulation of the GAA model with $\alpha=-0.8$, $\lambda=1$ and spatially uncorrelated hopping disorder of  varying strengths $\sigma=0,0.5$ and $1$ evolved up to time {$\tau=1{\rm{ms}}$} ($\tau\approx8.8{\hbar}/t_{max}$, where $t_{max}$ is the maximum hopping) starting from an initial population in the central site. Increased disorder leads to more localized states~\cite{roy2020interplay} which is reflected in the lesser spread of the simulated wavepacket.

The NPR values of the time-evolved wavepacket versus the strength $\lambda$ of the GAA potential (with hopping disorder) for different values of $\alpha$ are plotted in Fig.~\ref{fig: exp_NPR_RP_diff_alpha}. The plots also compare this with the ideal tight-binding model and MSL simulations incorporating the effect of experimental deviations. In the MSL simulations, a Gaussian wavepacket with width $\Delta p=0.07\hbar k$ is used to represent the initial state of the BEC, which is subject to an evolving sinusoidal potential modeling the counter-propagating laser beams. The width of the initial state in the momentum space estimated from the observed Rabi oscillations is $\Delta p = 0.07\hbar k$. The Hamiltonian at time $\tau$ is given by
\begin{equation}
    H(x,\tau)=-\frac{\hbar^{2}}{2m}\frac{\partial ^{2}}{\partial x^{2}} +\sum_{{i=-{n}/{2}}}^{{n}/{2}-1}\hbar\Omega_{i} \cos(2kx-\omega_{i}\tau)
    \label{Eq: Simulation Hamiltonian}
\end{equation}
where $\Omega_{i}$ is the angular Rabi frequency, $n$ is the number of tones or hopping bonds (20 in our case) and {$\omega_i{=}4E_{r}(2i+1)/\hbar+\delta_i$. $4E_{r}(2i+1)/\hbar$ is the resonant frequency between site $i$ and $i+1$. $E_r$ is the recoil energy, $E_r{=}\hbar^{2} k^{2}/2m$, and $\delta_i$ is the detuning chosen to match the quasiperiodic onsite energy difference between neighboring sites 
\begin{equation}
    \hbar\delta_i=\mu_{i+1}-\mu_{i}.
\end{equation}
The Rabi frequency is tuned to produce the right hopping strengths
\begin{equation}
\hbar\Omega_{i}/2=t_i.
\end{equation}
\subsection*{Numerical simulations of the MSL}
Numerical simulations of the MSL directly simulate Eq.~\ref{Eq: Simulation Hamiltonian} by a well-controlled approximation using a split operator method~\cite{fleck1978gpe}. The evolution operator $U(\Delta \tau)$, corresponding to a small time step $\Delta \tau$ under the Hamiltonian in Eq.~\ref{Eq: Simulation Hamiltonian} , is decomposed into its position- and momentum-space components, $U_x$ and $U_p$, respectively. Since each operator is diagonal in its corresponding basis, the wavefunction is alternately Fourier transformed between position and momentum space, allowing $U_x$ and $U_p$ to be applied successively to evolve the system (Also see Appendix~\ref{appendix: msl simulations}). The MSL simulation incorporates off-resonant excitations which effectively lead to deviations from the ideal tight-binding model corresponding to weak longer range hopping terms. 

Additionally, it accounts for finite momentum width effects of the BEC ($\Delta p=0.07\hbar k$). The BEC is a superposition of many momentum eigenstates centered around zero because of its momentum width. These components will evolve differently when subjected to the MSL due to the dependence of $\omega_i$ on the momentum components forming the lattice. The evolution will be similar at sufficiently small times, whereas at larger times they will deviate leading to a gradual loss of coherent interference. Experimental results agree well with the MSL simulations and qualitatively match the ideal tight-binding model, suggesting that off-resonant excitations and the momentum width of the initial state are main sources of deviation.

\section{Results\label{sec:Results}}

At $\alpha,\sigma{=}0$, the NPR is large in the delocalized phase($\lambda{<}2$), consistent with rapid wavepacket expansion. In contrast for $\lambda{>}2$, the NPR is small consistent with localization. Disorder increases localization for all $\lambda$ and $\alpha$.  With increasing  $\sigma$, the NPRs in the delocalized phase decrease, indicating localization albeit with large localization lengths. This is consistent with the expectation that the delocalization transition is turning into a crossover from weakly to strongly localized regimes.

The localizing effect of disorder is more pronounced at positive $\alpha$ as seen in the sharp decrease of wavefunction spread in the $\alpha=0.8$ case (Fig.~\ref{fig: exp_NPR_RP_diff_alpha}) in contrast to $\alpha=-0.8$. 
This behavior is a consequence of the choice of the site from which the wavepacket was released. Inset of the each panel shows the disorder profile. At $\alpha=0.8$, for the choice of the AA phase $\phi=0$ the central site has no nearby sites withe similar values of potentials. This absence of nearby resonances favors localization and suppressed tunneling of the initial packet into nearby sites. On the other hand at $\alpha=-0.8$ the opposite is true and resulting in sharp localization of the wavepacket setting in at a larger value of $\lambda$.
The $\alpha$ dependent features mentioned here are not universal and depends on the choice of the initial site and $\phi$ (or equivalently with the spectral weight of the initial state above and below the mobility edge). However even such non-universal microscopic features are faithfully captured by the MSL experiments. 

In Figure~\ref{fig: exp_NPR_RP_diff_alpha} we have studied the dynamics under a range of disorder width values. We find that the MSL dynamics in the cases with stronger disorder (larger $\sigma$) exhibit better agreement with the ideal dynamics. At small disorder the MSL exhibited slower wavepacket spread than the expected ideal simulation as revealed by the smaller NPRs in the MSL (in comparison with the ideal NPR values in the tight binding model) when the tunneling disorder width $\sigma$ was smaller. Numerical simulations of the MSL showed relatively better agreement with the MSL experiments in comparison with the ideal simulations.

The qualitative trends in the return probability also are captured faithfully by the MSL experiments. The return probability showed a close agreement with the numerical MSL simulations at all disorder values in the delocalized phase while the ideal tight binding simulation suggested slightly lower values of RP. In contrast in the localized regime, both numerical MSL and ideal simulations produced larger RP values in comparison with the MSL experiments. The deviations between the experimental data and the MSL simulations may arise from two physical effects not included in the model. Atomic collisions in the BEC can give rise to an effective interaction \cite{an2018Correlated}  while relative phase fluctuations between the two lattice beams can add noise both of these can result in deviations away from the MSL numerical simulations.

\begin{figure}[htbp]
    \centering
    \includegraphics[width=0.95\linewidth]{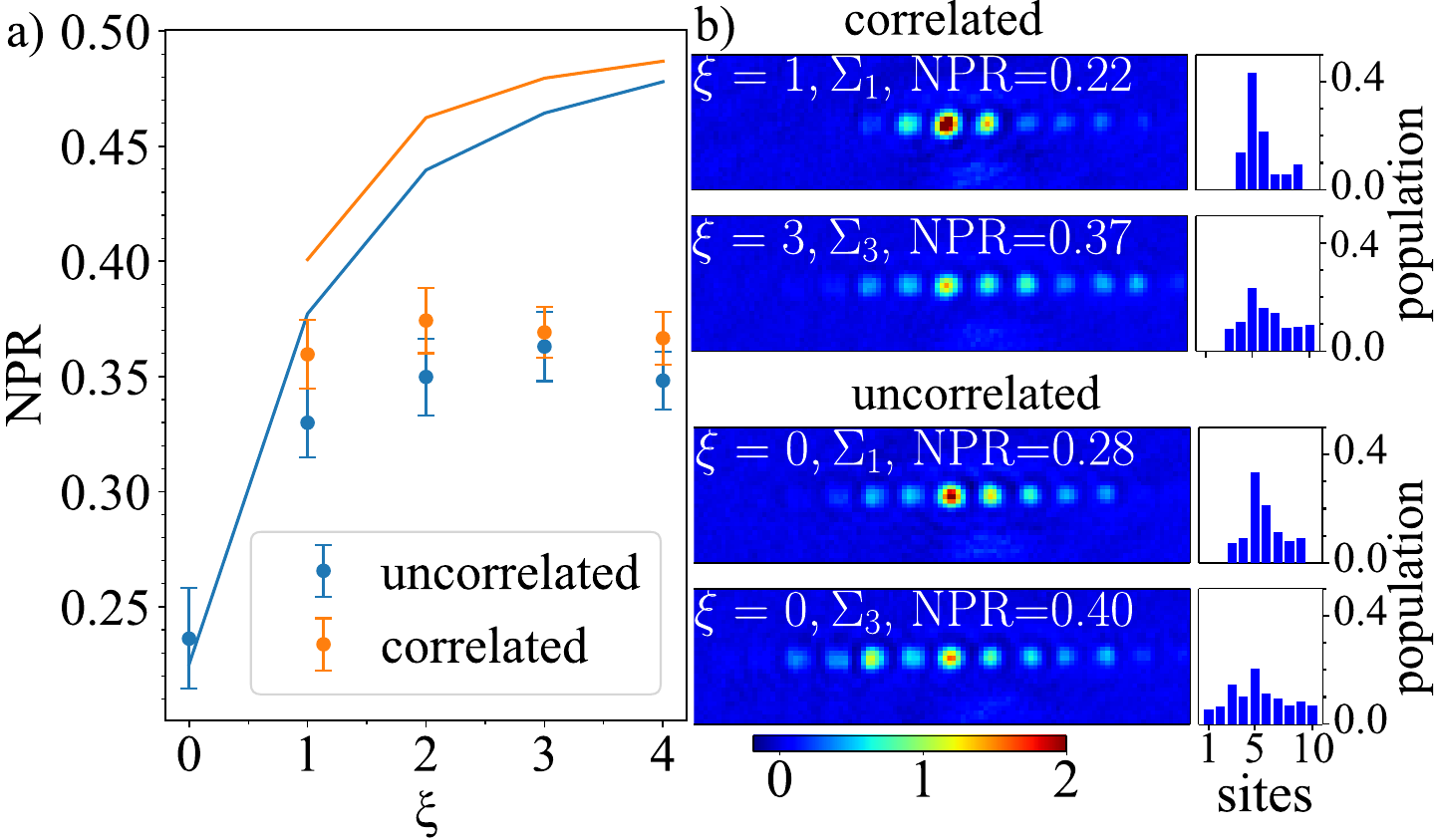}
        \caption{(a) NPR averaged over 50 realizations vs correlation length $\xi$ for $\alpha=0, \lambda=1$ and initial $\sigma=1$. For each $\xi$, the correlated hopping disorder is compared with the uncorrelated hopping disorder having the same typical standard deviation (averaged over 50 realizations) $\Sigma_\xi$; $\Sigma_1=0.27$, $\Sigma_2=0.16$, $\Sigma_3=0.11$, $\Sigma_4=0.08$. $\xi=0$ is simply the uncorrelated case. Solid lines show MSL simulations and filled circles show experimental measurements. (b) The optical density plots and population vs position (right) for a specific realization.\label{fig: correlated_disorder_and_ipr}}
\end{figure}

Lastly, we investigate the experimental realization of spatially correlated disorder in hopping strengths of the AA model. For each $\xi$, we perform the experiment for 50 disorder realizations and introduce spatial correlations by Gaussian smoothing (Eq.~\ref{eq:gaussian_smoothing}) of the hopping terms. Figure~\ref{fig: correlated_disorder_and_ipr} shows the NPR of the time-evolved state as a function of the disorder correlation length \(\xi\). 

We also show a comparison with the corresponding uncorrelated disorder in Fig.~\ref{fig: correlated_disorder_and_ipr}. Experiments are done in system with tunneling disorder added to AA model in the delocalized phase. 
Smoothening used to generate disorder correlations can reduce the variance of the disorder distribution. For instance an uncorrelated disorder sample of std. deviation $\Sigma_{0}\approx 0.58$ can shrink to $\Sigma_{1}\approx 0.27$ at $\xi=1$ and $\Sigma_{4}\approx 0.08$ at $\xi=4$. This can result in faster wavefunction spread. This explains the increasing NPR with $\xi$ seen in Fig.~\ref{fig: correlated_disorder_and_ipr}. To separate this effect when comparing correlated and uncorrelated disorder, we need to compare the NPR for the smoothened disorder with the NPR for the uncorrelated disorder after rescaling the width of the latter such that the standard deviation of both disorder distributions match. Note that rescaling does not alter the correlation length $\xi$.

The disorder-averaged NPR is consistently lower for the uncorrelated case compared to the correlated one. The disorder width dependence as well as the correlation dependence are consistent with expectations based on the results presented in Sec.~\ref{sec:model} where numerical simulation of the spread as well as transmission based localization length estimates suggested more wavefunction spread with decreasing tunneling disorder strengths and increasing correlation scale $\xi$. These trends are captured correctly in the experimental data. 

Numerical simulations of the MSL under the corresponding correlated and uncorrelated disorder are also shown in Fig.~\ref{fig: correlated_disorder_and_ipr}. Similar to what was observed in Fig.~\ref{fig: exp_NPR_RP_diff_alpha} at weak disorder (which appears at larger $\xi$), the numerical MSL simulation shows a larger deviation from the MSL experiments. Even though the numerical NPR does not match quantitatively, both numerical MSL simulation and experiments in the MSL experiments show slightly faster spread of the wavefunction in the presence of the disorder correlations.

\section{Conclusions\label{sec:Summary}}
In this work, we have presented an experimental realization of the GAA model with added hopping disorder in a MSL with an $^{87}$Rb BEC. {The addition of hopping disorder suppresses transport and increases localization across the parameter regime explored here.} It also smears out the transition between localized and delocalized phases. NPR dynamics agrees qualitatively with the ideal tight-binding model, but we achieve much better quantitative agreement once important sources of experimental deviations, namely off-resonant excitations and finite initial-state widths, are incorporated. Experimental fidelity can thus be improved by minimizing off-resonant excitations through pulse shaping. Delta-kick collimation can be employed to reduce the momentum spread of the BEC. In the correlated disorder regime, partially delocalized eigenstates emerge over the correlation length, and future experiments with engineered initial states could selectively probe these features. MSL thus provides a versatile platform with site-resolved control over hopping and on-site energies, getting us closer to high-fidelity simulations of quantum transport beyond quasiperiodic models.

 \section*{acknowledgements}
 All the authors thank the I-HUB quantum technology foundation (I-HUB QTF) at IISER Pune for support. J.M.S. and R.A.S. acknowledge research fellowship from Council of Scientific and Industrial Research (CSIR), Government of India. M.B. acknowledges funding from Qmet PARIMANA Fellowship, an initiative of DST National Quantum Mission. S.R acknowledges funding from the Prime Minister's Research Fellowship (PMRF).

\clearpage
\appendix
\begin{figure*}
\includegraphics[width=\textwidth]{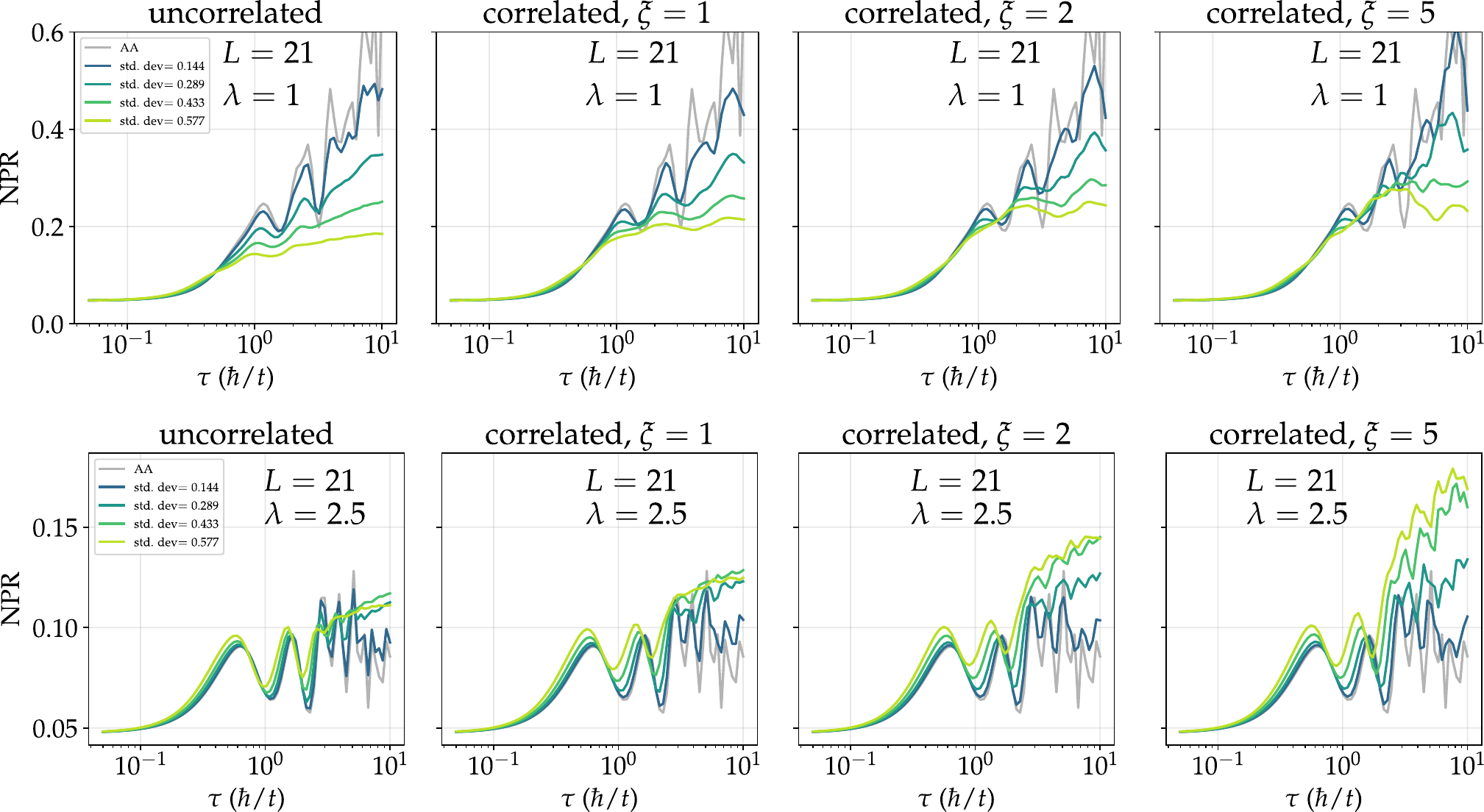}
\caption{Similar to Fig.~\ref{fig:NPRDynamics} but for $L=21$. The results are qualitatively the same as in the $L=300$ case. \label{fig:NPRDynamics21}}
\end{figure*}

\section{Simulations of the momentum space lattice}
\label{appendix: msl simulations}

\begin{figure*}[ht!]
    \centering
    \includegraphics[width=0.9\linewidth]{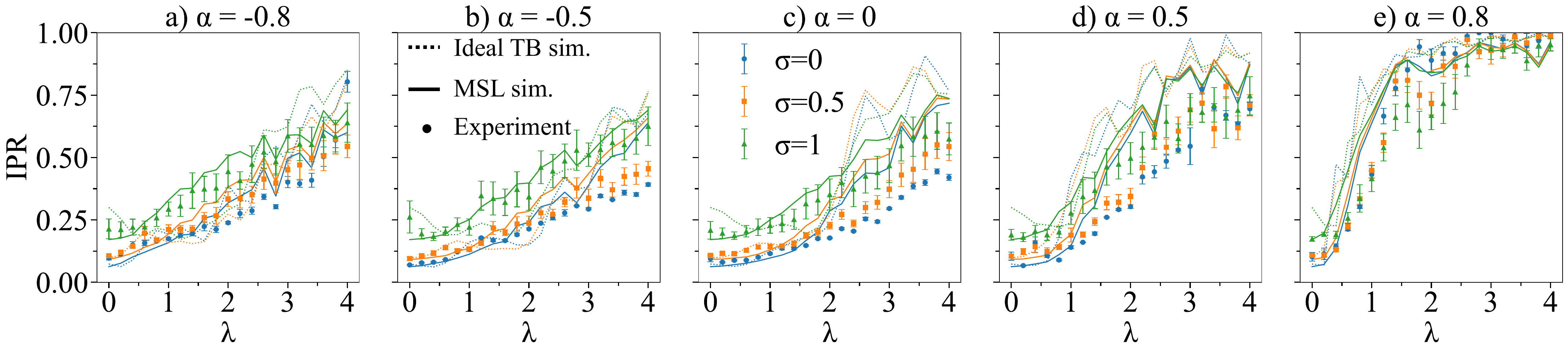}
    \caption{The IPR vs lambda plots for the GAA model. The IPR gives a clearer picture of the effect of adding disorder at higher $\lambda$ values than the NPR. The solid lines are simulations of the MSL experiment, the dotted lines are ideal tight-binding (TB) simulations,  and the markers with error bar (standard error) are experimental data.}
    \label{fig:IPR_all_alphas}
\end{figure*}

We perform numerical simulations of the ideal tight-binding model as well as a more complete simulation of the BEC dynamics in the momentum space lattice. The simulation of the BEC dynamics takes into account the momentum width of the BEC, off-resonant excitations of the lattice beams, and loss of coherent dynamics due to the momentum states splitting apart. The Hamiltonian at time $\tau$ is given by:
\begin{equation}
    {H}(x,\tau)=-\frac{\hbar^{2}}{2m}\frac{\partial^{2} }{\partial x^{2}} +\sum_{i=\frac{-n}{2}}^{\frac{n}{2}-1}\hbar\Omega_{i} \cos(2kx-\omega_{i}\tau)
    \label{Eq: Simulation Hamiltonian}
\end{equation}
where, $\Omega_{i}$ is the angular Rabi frequency. n is the number of tones (which in our case is 20) and $\omega_i{=}4E_{r}(2i+1)/\hbar+\delta_i$. $4E_{r}(2i+1)/\hbar$ is the resonant frequency between site $i$ and $i+1$. $E_r$ is the recoil energy, $E_r{=}\hbar^{2} k^{2}/2m$, and $\delta_i$ is the detuning. We use the split-operator method which is detailed in \cite{fleck1978gpe}. The evolution operator which can be approximated as the propagator $U(\tau,\Delta \tau)=\exp[-iH(\tau)\Delta \tau/\hbar]$. It is split into momentum and position space components $U_p(\Delta \tau)$ and $U_{x}(\tau, \Delta \tau)$. Where, $U_p(\Delta \tau)=\exp[-ip^2\Delta \tau/2m\hbar]$ and $U_x(\tau, \Delta \tau)=\exp[-iV(x,\tau) \Delta \tau/\hbar]$. From Eq. \ref{Eq: Simulation Hamiltonian}, we get $V(x,\tau)=\sum_{i=-\frac{n}{2}}^{\frac{n}{2}-1}\hbar\Omega_{i}\cos(2kx-\omega_{i}\tau)$ for the Momentum Space Lattice (MSL). Using Strang splitting the propagator can be decomposed as:
\begin{multline}
    U(\tau, \Delta \tau)=U_p(\Delta \tau/2)U_x(\tau+\Delta \tau/2, \Delta \tau)U_p(\Delta \tau/2)\\+O((\Delta \tau)^3)
\end{multline}
The second term quantifies the error from this decomposition. For an evolution of N steps of $\Delta \tau$ starting from $\tau=0$, the propagator becomes:
\begin{multline}
    U_N(0, \Delta \tau)= \mathcal{T}\prod_{k=0}^{N-1}U(k\Delta \tau,\Delta \tau) +O((\Delta \tau)^2)
\end{multline}
The wavefunction is transformed alternately between the position and momentum space using the Discrete Fourier Transform (DFT) and the respective evolution operators in each space are applied sequentially.\\ 
We take experimental data for the Rabi-oscillations between the 0th and 1st site. We can then approximately calculate the momentum width of the BEC by fitting the damped Rabi oscillations. We get a momentum width of $\Delta p=0.07\hbar k$. So we initialize the BEC as a Gaussian with momentum width $\Delta p=0.07\hbar k$ for all simulations of the GAA model.\\

\section{Inverse participation ratio}

The IPR is calculated from the measured site-resolved populations $\mathcal{P}_i = |\langle i|\psi(\tau)\rangle|^2$ as
\begin{equation}
\mathrm{IPR}(\tau) = {\scriptstyle{\sum}}_i \mathcal{P}_i^2
\end{equation}
providing a dynamical probe of the underlying localization \cite{schreiber2015observation,bordia2017periodically}. 
It is related to the NPR by
\begin{equation}
  \mathrm{IPR(\tau)} = \frac{1}{L \,(\mathrm{NPR}(\tau))},
  \label{eq:npr}
\end{equation}
where $L$ is the system size. The IPR of the eigenstates is $O(1/\chi)$ for states localized over $\chi$ sites and is $O(1/\mathrm{L})$ for extended states, providing a measure of spatial localization. Importantly, upon introducing disorder, the effect of localization becomes clearly visible in the IPR even at large $\lambda$, where the NPR remains close to zero and exhibits only weak variation across disorder strengths.

\section{Transmission coefficient and estimation of localization length\label{sec:transmission}}
In this section, we summarize the calculation of the transmission coefficients which was used in the calculation of the localization lengths in Fig.~\ref{fig:transmission}. The system with the Hamiltonian $H$ is coupled to a semi infinite tight binding chain (with no onsite potential and with a bandwidth larger than $H$) on either side acting as leads. The retarded surface Greens function of the leads on the site adjacent to the system can be estimated analytically for the semi-infinite chain
\begin{equation}
  \mathfrak{g}(E) =
    \dfrac{E - i\sqrt{4t_L^2 - E^2}}{2t_L^2},  \quad (|E| < 2t_L),
\end{equation}
The transmission coefficient $T(E)$ is proportional to $|G_{1L}(E)|^2$ where $G_{ij}$ is the matrix element of the retarded Green's function between the $i$ and $j$ sites.  $1$ and $L$ are the first and last sites of the system. $G_{1L}(E)$ can be calculated as 
\begin{equation}
G_{1L} = G_{LL}\prod_{i=1}^{L-1} g_i t_i
\end{equation}
where $g_{n}$ for $n=1,2\dots L-1$ is the Green's function of a chain of length $n$ coupled to the semi infinite lead only on the left. This function $g_n(E)$ can be calculated recursively. In the discussion below $\Sigma=|t_{\rm LS}|^2 \mathfrak{g} $ ($t_{\rm LS}$ is the lead system coupling taken to be $1$) is the lead self energy. Since we use identical leads, we do not need to distinguish the their self-energy functions. The recursion relations are 
\begin{align}
&g_1= \frac{1}{E - \mu_1 - \Sigma(E)}\nonumber\\
&g_n= \frac{1}{E - \mu_n - t_{n-1}^2\, g_{n-1}} , {\text{ for }} n = 2, 3, \dots L-1\nonumber
\end{align}
where $\mu_i$ are the onsite potentials and $t_{n}$ is the hopping strength between $n$th and $(n+1)$th sites. Lastly, $G_{LL}$ is given by 
\begin{equation}
  G_{LL} = \frac{1}{E - \mu_L - t_{L-1}^2 g_{L-1} - \Sigma} .
\end{equation}
where the effect of the right lead appears through $\Sigma$.
The $L$ dependence of $T$ is contained in $|G_{1L}|^2$ and we expect it to behave as $e^{-L/\xi_{\rm loc}}$ for long enough chains in the localized phase. $\ln T$ is a self-averaging quantity in 1D. We therefore estimate $ 1/{\xi_{\rm loc}} $ as the slope of $\langle \ln T(E)\rangle$ vs $L$ where the average is over disorder realizations.
\bibliography{ref}

@article{anderson1958absence,
  title={Absence of diffusion in certain random lattices},
  author={Anderson, Philip W},
  journal={Physical Review},
  volume={109},
  number={5},
  pages={1492},
  year={1958},
  publisher={APS}
}

@article{evers2008anderson,
  title={Anderson transitions},
  author={Evers, Ferdinand and Mirlin, Alexander D},
  journal={Reviews of Modern Physics},
  volume={80},
  number={4},
  pages={1355--1417},
  year={2008},
  publisher={APS}
}

@article{lagendijk2009fifty,
  title={Fifty years of {Anderson} localization},
  author={Lagendijk, Ad and Tiggelen, Bart van and Wiersma, Diederik S},
  journal={Physics today},
  volume={62},
  number={8},
  pages={24--29},
  year={2009},
  publisher={AIP Publishing}
}

@article{Eggarter1978,
  title = {Singular behavior of tight-binding chains with off-diagonal disorder},
  author = {Eggarter, T. P. and Riedinger, R.},
  journal = {Phys. Rev. B},
  volume = {18},
  issue = {2},
  pages = {569--575},
  numpages = {0},
  year = {1978},
  month = {Jul},
  publisher = {American Physical Society},
  doi = {10.1103/PhysRevB.18.569},
  url = {https://link.aps.org/doi/10.1103/PhysRevB.18.569}
}

@article{aubry1980analyticity,
  title={Analyticity breaking and {Anderson} localization in incommensurate lattices},
  author={Aubry, Serge and Andr{\'e}, Gilles},
  journal={Ann. Israel Phys. Soc},
  volume={3},
  number={133},
  pages={18},
  year={1980}
}

@article{billy2008direct,
  title={Direct observation of {Anderson} localization of matter waves in a controlled disorder},
  author={Billy, Juliette and Josse, Vincent and Zuo, Zhanchun and Bernard, Alain and Hambrecht, Ben and Lugan, Pierre and Cl{\'e}ment, David and Sanchez-Palencia, Laurent and Bouyer, Philippe and Aspect, Alain},
  journal={Nature},
  volume={453},
  number={7197},
  pages={891--894},
  year={2008},
  publisher={Nature Publishing Group}
}

@article{roati2008anderson,
  title={Anderson localization of a non-interacting {Bose--Einstein} condensate},
  author={Roati, Giacomo and D’Errico, Chiara and Fallani, Leonardo and Fattori, Marco and Fort, Chiara and Zaccanti, Matteo and Modugno, Giovanni and Modugno, Michele and Inguscio, Massimo},
  journal={Nature},
  volume={453},
  number={7197},
  pages={895--898},
  year={2008},
  publisher={Nature Publishing Group UK London}
}

@article{kondov2015disorder,
  title={Disorder-induced localization in a strongly correlated atomic {Hubbard} gas},
  author={Kondov, SS and McGehee, WR and Xu, W and DeMarco, B},
  journal={Physical Review Letters},
  volume={114},
  number={8},
  pages={083002},
  year={2015},
  publisher={APS}
}

@article{semeghini2015measurement,
  title={Measurement of the mobility edge for 3D {Anderson} localization},
  author={Semeghini, Giulia and Landini, Manuele and Castilho, Patricia and Roy, Sanjukta and Spagnolli, Giacomo and Trenkwalder, Andreas and Fattori, Marco and Inguscio, Massimo and Modugno, Giovanni},
  journal={Nature Physics},
  volume={11},
  number={7},
  pages={554--559},
  year={2015},
  publisher={Nature Publishing Group UK London}
}

@article{schwartz2007transport,
  title={Transport and {Anderson} localization in disordered two-dimensional photonic lattices},
  author={Schwartz, Tal and Bartal, Guy and Fishman, Shmuel and Segev, Mordechai},
  journal={Nature},
  volume={446},
  number={7131},
  pages={52--55},
  year={2007},
  publisher={Nature Publishing Group UK London}
}

@article{chabanov2000statistical,
  title={Statistical signatures of photon localization},
  author={Chabanov, Azriel A and Stoytchev, Moshe and Genack, Azriel Z},
  journal={Nature},
  volume={404},
  number={6780},
  pages={850--853},
  year={2000},
  publisher={Nature Publishing Group}
}

@article{lahini2009observation,
  title={Observation of a localization transition in quasiperiodic photonic lattices},
  author={Lahini, Yoav and Pugatch, Rami and Pozzi, Francesca and Sorel, Marc and Morandotti, Roberto and Davidson, Nir and Silberberg, Yaron},
  journal={Physical Review Letters},
  volume={103},
  number={1},
  pages={013901},
  year={2009},
  publisher={APS}
}

@article{wiersma1997localization,
  title={Localization of light in a disordered medium},
  author={Wiersma, Diederik S and Bartolini, Paolo and Lagendijk, Ad and Righini, Roberto},
  journal={Nature},
  volume={390},
  number={6661},
  pages={671--673},
  year={1997},
  publisher={Nature Publishing Group UK London}
}

@article{kraus2012topological,
  title={Topological states and adiabatic pumping in quasicrystals},
  author={Kraus, Yaacov E and Lahini, Yoav and Ringel, Zohar and Verbin, Mor and Zilberberg, Oded},
  journal={Physical Review Letters},
  volume={109},
  number={10},
  pages={106402},
  year={2012},
  publisher={APS}
}

@article{verbin2013observation,
  title={Observation of topological phase transitions in photonic quasicrystals},
  author={Verbin, Mor and Zilberberg, Oded and Kraus, Yaacov E and Lahini, Yoav and Silberberg, Yaron},
  journal={Physical Review Letters},
  volume={110},
  number={7},
  pages={076403},
  year={2013},
  publisher={APS}
}

@article{guo2021stark,
  title = {Stark Many-body Localization on a Superconducting Quantum Processor},
  author = {Guo, Qiujiang and Cheng, Chen and Li, Hekang and Xu, Shibo and Zhang, Pengfei and Wang, Zhen and Song, Chao and Liu, Wuxin and Ren, Wenhui and Dong, Hang and Mondaini, Rubem and Wang, H.},
  journal = {Phys. Rev. Lett.},
  volume = {127},
  issue = {24},
  pages = {240502},
  numpages = {7},
  year = {2021},
  month = {Dec},
  publisher = {American Physical Society},
  doi = {10.1103/PhysRevLett.127.240502},
  url = {https://link.aps.org/doi/10.1103/PhysRevLett.127.240502}
}

@article{luschen2018SPME,
  title={Single-particle mobility edge in a one-dimensional quasiperiodic optical lattice},
  author={L{\"u}schen, Henrik P and Scherg, Sebastian and Kohlert, Thomas and Schreiber, Michael and Bordia, Pranjal and Li, Xiao and Sarkar, S Das and Bloch, Immanuel},
  journal={Physical Review Letters},
  volume={120},
  number={16},
  pages={160404},
  year={2018},
  publisher={APS}
}

@article{bordia2017periodically,
  title={Periodically driving a many-body localized quantum system},
  author={Bordia, Pranjal and L{\"u}schen, Henrik P and Hodgman, Sean S and Schreiber, Michael and Bloch, Immanuel and Schneider, Ulrich},
  journal={Nature Physics},
  volume={13},
  number={5},
  pages={460--464},
  year={2017},
  publisher={Nature Publishing Group}
}

@article{schreiber2015observation,
  title={Observation of many-body localization of interacting fermions in a quasirandom optical lattice},
  author={Schreiber, Michael and Hodgman, Sean S and Bordia, Pranjal and L{\"u}schen, Henrik P and Fischer, Mark H and Vosk, Ronen and Altman, Ehud and Schneider, Ulrich and Bloch, Immanuel},
  journal={Science},
  volume={349},
  number={6250},
  pages={842--845},
  year={2015},
  publisher={American Association for the Advancement of Science}
}

@article{biddle2010predicted,
   title={Predicted Mobility Edges in One-Dimensional Incommensurate Optical Lattices: An Exactly Solvable Model of {Anderson} Localization},
   volume={104},
   ISSN={1079-7114},
   url={http://dx.doi.org/10.1103/PhysRevLett.104.070601},
   DOI={10.1103/physrevlett.104.070601},
   number={7},
   journal={Physical Review Letters},
   publisher={American Physical Society (APS)},
   author={Biddle, J. and Das Sarma, S.},
   year={2010},
   month=feb }

@article{li2017mobility,
  title={Mobility edges in one-dimensional bichromatic incommensurate potentials},
  author={Li, Xiao and Li, Xiaopeng and Das Sarma, S},
  journal={Physical Review B},
  volume={96},
  number={8},
  pages={085119},
  year={2017},
  publisher={APS}
}

@article{li2015many,
  title={Many-body localization and quantum nonergodicity in a model with a single-particle mobility edge},
  author={Li, Xiaopeng and Ganeshan, Sriram and Pixley, JH and Das Sarma, S},
  journal={Physical Review Letters},
  volume={115},
  number={18},
  pages={186601},
  year={2015},
  publisher={APS}
}

@article{roy2018multifractality,
  title={Multifractality without fine-tuning in a {Floquet} quasiperiodic chain},
  author={Roy, Sthitadhi and Khaymovich, Ivan M and Das, Arnab and Moessner, Roderich},
  journal={SciPost Physics},
  volume={4},
  number={5},
  pages={025},
  year={2018}
}

@article{torres2017dynamical,
  title={Dynamical manifestations of quantum chaos: correlation hole and bulge},
  author={Torres-Herrera, EJ and Santos, Lea F},
  journal={Philosophical Transactions of the Royal Society A: Mathematical, Physical and Engineering Sciences},
  volume={375},
  number={2108},
  pages={20160434},
  year={2017},
  publisher={The Royal Society Publishing}
}

@article{ganeshan2015nearest,
  title={Nearest neighbor tight binding models with an exact mobility edge in one dimension},
  author={Ganeshan, Sriram and Pixley, JH and Das Sarma, S},
  journal={Physical Review Letters},
  volume={114},
  number={14},
  pages={146601},
  year={2015},
  publisher={APS}
}

@article{abrahams1979scaling,
  title={Scaling theory of localization: {Absence} of quantum diffusion in two dimensions},
  author={Abrahams, Elihu and Anderson, Philip W and Licciardello, Donald C and Ramakrishnan, Tiruppattur V},
  journal={Physical Review Letters},
  volume={42},
  number={10},
  pages={673},
  year={1979},
  publisher={APS}}

@article{an2021interactions,
  title={Interactions and mobility edges: {Observing} the generalized {A}ubry-{A}ndr{\'e} model},
  author={An, Fangzhao Alex and Padavi{\'c}, Karmela and Meier, Eric J and Hegde, Suraj and Ganeshan, Sriram and Pixley, JH and Vishveshwara, Smitha and Gadway, Bryce},
  journal={Physical Review Letters},
  volume={126},
  number={4},
  pages={040603},
  year={2021},
  publisher={APS}
}

@article{harper1955single,
  title={Single band motion of conduction electrons in a uniform magnetic field},
  author={Harper, Philip George},
  journal={Proceedings of the Physical Society. Section A},
  volume={68},
  number={10},
  pages={874},
  year={1955},
  publisher={IOP Publishing}
}

@article{roy2020interplay,
	author = {Roy, Shilpi and Basu, Saurabh},
	title = {Interplay of off-diagonal random disorder and quasiperiodic potential in a one-dimensional {Aubry-André} model},
	DOI= "10.1209/0295-5075/128/47005",
	url= "https://doi.org/10.1209/0295-5075/128/47005",
	journal = {EPL},
	year = 2019,
	volume = 128,
	number = 4,
	pages = "47005",
}

@article{fleck1978gpe,
  title={Time-dependent propagation of high-energy laser beams through the atmosphere: II},
  author={Joseph A. Fleck and J R Morris and Michael D. Feit},
  journal={Applied physics},
  year={1978},
  volume={14},
  pages={99-115},
  url={https://api.semanticscholar.org/CorpusID:38275310}
}

@article{meier2016simulator,
  title = {Atom-optics simulator of lattice transport phenomena},
  author = {Meier, Eric J. and An, Fangzhao Alex and Gadway, Bryce},
  journal = {Phys. Rev. A},
  volume = {93},
  issue = {5},
  pages = {051602},
  numpages = {5},
  year = {2016},
  month = {May},
  publisher = {American Physical Society},
  doi = {10.1103/PhysRevA.93.051602},
  url = {https://link.aps.org/doi/10.1103/PhysRevA.93.051602}
}

@article{li2023observation,
  title = {Observation of critical phase transition in a generalized {Aubry-André-Harper} model with superconducting circuits},
  volume = {9},
  ISSN = {2056-6387},
  url = {http://dx.doi.org/10.1038/s41534-023-00712-w},
  DOI = {10.1038/s41534-023-00712-w},
  number = {1},
  journal = {npj Quantum Information},
  publisher = {Springer Science and Business Media LLC},
  author = {Li,  Hao and Wang,  Yong-Yi and Shi,  Yun-Hao and Huang,  Kaixuan and Song,  Xiaohui and Liang,  Gui-Han and Mei,  Zheng-Yang and Zhou,  Bozhen and Zhang,  He and Zhang,  Jia-Chi and Chen,  Shu and Zhao,  S. P. and Tian,  Ye and Yang,  Zhan-Ying and Xiang,  Zhongcheng and Xu,  Kai and Zheng,  Dongning and Fan,  Heng},
  year = {2023},
  month = apr 
}

@article{shimasaki2024anomalous,
  title = {Anomalous localization in a kicked quasicrystal},
  volume = {20},
  ISSN = {1745-2481},
  url = {http://dx.doi.org/10.1038/s41567-023-02329-4},
  DOI = {10.1038/s41567-023-02329-4},
  number = {3},
  journal = {Nature Physics},
  publisher = {Springer Science and Business Media LLC},
  author = {Shimasaki,  Toshihiko and Prichard,  Max and Kondakci,  H. Esat and Pagett,  Jared E. and Bai,  Yifei and Dotti,  Peter and Cao,  Alec and Dardia,  Anna R. and Lu,  Tsung-Cheng and Grover,  Tarun and Weld,  David M.},
  year = {2024},
  month = jan,
  pages = {409–414}
}

@article{an2018Correlated,
  title = {Correlated Dynamics in a Synthetic Lattice of Momentum States},
  author = {An, Fangzhao Alex and Meier, Eric J. and Ang'ong'a, Jackson and Gadway, Bryce},
  journal = {Phys. Rev. Lett.},
  volume = {120},
  issue = {4},
  pages = {040407},
  numpages = {6},
  year = {2018},
  month = {Jan},
  publisher = {American Physical Society},
  doi = {10.1103/PhysRevLett.120.040407},
  url = {https://link.aps.org/doi/10.1103/PhysRevLett.120.040407}
}
\end{document}